\documentstyle[12pt]{article}
\newcommand{\be}[1]{\begin{equation} \label{(#1)}}
\newcommand{\ee}{\end{equation}}
\newcommand{\ba}[1]{\begin{eqnarray} \label{(#1)}}
\newcommand{\ea}{\end{eqnarray}}
\newcommand{\nn}{\nonumber}

\newcommand{\rf}[1]{(\ref{(#1)})}

\begin{document}

\bigskip

\begin{center}

{\bf On  SUSY Dark Matter Detection with \\
Spinless Nuclei\footnotemark[1].}

\bigskip

{V.A. Bednyakov, H.V. Klapdor-Kleingrothaus$^*$, S.G. Kovalenko
\bigskip

{\it Joint Institute for Nuclear Research, Dubna, Russia}
\bigskip

$^*${\it
Max-Plank-Institut f\"{u}r Kernphysik, D-6900, Heidelberg,
Germany}

}
\end{center}

\begin{abstract}
        We investigate the role of nuclear spin in elastic scattering of
        Dark Matter (DM) neutralinos from nuclei in the framework of
        the Minimal SUSY standard model (MSSM).
        The relative contribution of spin-dependent axial-vector and
        spin-independent scalar interactions to the  event rate in a
        DM detector has been analyzed for various nuclei.
        Within general assumptions about the nuclear and nucleon structure
        we find that for nuclei with atomic weights $A > 50$ the
        spin-independent part of the event rate $R_{si}$ is larger than the
        spin-dependent one $R_{sd}$ in the domain of the MSSM parameter
        space allowed by the known experimental data and where  the
        additional constraint for the total event rate
        $R =R_{sd} + R_{si} > 0.01$ is satisfied.
        The latter reflects realistic sensitivities of present and
        near future DM detectors.
        Therefore we expect equal chances for discovering the DM event
        either with spin-zero or with spin-non-zero isotopes if their
        atomic weights are $A_{1} \sim A_{2} > 50$.
\end{abstract}
\bigskip

\footnotetext[1]{
This work was supported in part by the Russian Foundation for Fundamental
Research (93-02-3744).
}
\newpage

\section{Introduction}
        Analysis of the data on distribution and motion of astronomical
        objects within our galaxy and far beyond indicates presence
        of a large amount of non-luminous dark matter (DM).
        According to estimations, it constitutes more than $90\%$ of the
        total mass of the universe  if  a
        mass density $\rho$ of the universe close to the critical value
        $ \rho_{crit}$ is  assumed.
        The exact equality
        $\Omega=\rho/\rho_{crit}=1$, corresponding to a flat universe,
        is supported by naturalness arguments and by inflation scenarios.
        Also, in our galaxy most of the mass should be in a dark halo.
        Detailed models predict a spherical form for the galaxy halo
        and a Maxwellian distribution for DM particle velocities
        in the galactic frame.
        The mass density of DM in the Solar system should be
        about $\rho \approx 0.3 $GeV$\cdot$cm$^{-3}$ and
        the DM particles should arrive at the earth's surface with mean
        velocities $v\approx$ 320 km/sec, producing a substantial flux
        $\Phi = \rho\cdot v/M$ ($\Phi>10^{7}$cm$^{-2}$ sec$^{-1}$
        for the particle mass M$\sim$ 1 GeV).
        Therefore one may hope to detect DM particles directly, for instance
        through the elastic scattering from nuclei inside a detector.

        The theory of primordial nucleosynthesis restricts the amount
        of baryonic matter in the universe to 10\%.
        Thus a dominant component of DM is non-baryonic.
        The recent data by the COBE satellite \cite{COBE} on  anisotropy
        in the cosmic background radiation and the theory of the
        formation of large scale structures
        of the universe lead to the conclusion that non-baryonic DM itself
        consists of a dominant ($70\%$) "cold" DM (CDM) and smaller
        ($30\%$) "hot" DM (HDM) component \cite{Taylor}, \cite{Davis}.

        The neutralino ($\chi$) is a favorable candidate for CDM.
        This is a Majorana ($\chi^{c}=\chi$) spin-half particle predicted
        by supersymmetric (SUSY) models.

        There are four neutralinos in the minimal SUSY extension of the
        standard model (MSSM).
        They are a mixture of gauginos ($\tilde{W}_{3}, \tilde{B})$
        and Higgsinos ($\tilde{H}_{1,2}$), which are SUSY partners of gauge
        ($W_{3}, B$) and Higgs ($H_{1,2}$) bosons.
        The DM neutralino $\chi$ is the lightest of them.
        Moreover, $\chi$ is assumed to be the lightest SUSY particle (LSP)
        which is stable in SUSY models with $R$-parity conservation.

        The problem of direct detection of the DM neutralino $\chi$  via
        elastic scattering off nuclei has been considered by many authors
        and remains a field of great experimental and theoretical
        activity \cite{Witt}-\cite{Bot1}.

        The final goal of theoretical calculations in this problem is the
        event rate $R$ for elastic $\chi$-nucleus scattering.
        In general, the spin-dependent ($R_{sd}$)
        and spin-independent ($R_{si}$) neutralino-nucleus interactions
        contribute to the event rate: $R=R_{sd}+R_{si}$.
        $R_{sd}$ vanishes for spinless nuclei and this fact is often
        regarded as a reason to assert spinless nuclei to be irrelevant
        for the DM neutralino detection as giving a much smaller event rate.
        One can meet this statement in the literature.
        However, this is right only if the spin-dependent interaction
        dominates in elastic neutralino scattering off nuclei with non-zero
        spin.

        In this paper we address the question on the role of nuclear
        spin in the DM neutralino detection.
        We investigate this problem in the framework of the MSSM.
        We avoid using specific nuclear and nucleon structure models but
        rather base our consideration on the known experimental data about
        nuclei and nucleon.
        It allows us to free the consideration of theoretical uncertainties
        specific for the structure models.
        To restrict the MSSM parameter space we use experimental constraints
        on SUSY-particle masses, the cosmological bound on neutralino
        relic abundance and the proton life-time constraint.

        We have found that $R_{si}$ contribution dominates in the total
        event rate $R$ for nuclei with atomic weight $A>50$ in the
        region of the MSSM parameter space where
        $R =R_{sd}+R_{si}< 0.01$.
        The lower bound  $0.01$ is far below the sensitivity of
        realistic present and near future DM detectors.
        Therefore we can exclude the region where  $R < 0.01$ as
        invisible for these detectors.

        We do {\it not\ } expect a crucial dependence of the
        DM event rate on the nuclear spin for detectors with target nuclei
        having an atomic weight larger than $50$.
        As a result, we expect equal chances for J~=~0 and J~$\neq$~0
        detectors to discover DM events.
        In particular, this conclusion supports the idea that presently
        operating $\beta\beta$-detectors with spinless nuclear target
        material can be successfully used for DM neutralino search.
        These highly developed set-ups (for a review see~\cite{Klapdor}),
        operating under extremely low background conditions, use detection
        technology which is suitable for the DM search.

\section{General Properties of the Neutralino - Nucleus Interactions.}

        A DM event is elastic neutralino-nucleus scattering causes the
        nuclear recoil detected by a detector.
        The event rate per unit mass of the target material depends on
        the distribution of the DM neutralinos in the solar vicinity and
        the cross section $\sigma_{el}(\chi A)$ of  neutralino-nucleus
        elastic scattering.
        One can calculate $\sigma_{el}(\chi A)$  starting from the
        neutralino-quark effective Lagrangian.
        In the most general form it can be given by the formula
\be{Lagr} 
  L_{eff} = \sum_{q}^{} \{ {\cal B}(q) -
        {\cal A}(q)\}\cdot\bar\chi\gamma_\mu\gamma_5\chi\cdot
                \bar q\gamma^\mu\gamma_5 q +
    \frac{m_q}{M_{W}} \cdot{\cal C}_{q}\cdot\bar\chi\chi\cdot\bar q q,
\ee
        where terms with the vector and pseudoscalar quark currents are
        omitted being negligible in the case of the non-relativistic
        DM neutralino with typical velocities $v_\chi\approx 10^{-3} c$.
        Following the tradition we retain in the first term the difference
        of two coefficients representing just one independent parameter of
        the Lagrangian.
        The coefficients ${\cal B}(q), {\cal A}(q), {\cal C}(q)$ depend on
        the SUSY model and will be considered in the next section.
        Here we survey general properties of $\chi$-$A$ scattering following
        from the Lagrangian \rf{Lagr}.

        To calculate $\sigma_{el}(\chi A)$ one should average
        the $\chi$-$q$ interactions sequentially over the nucleon and
        the nuclear structure.
        The first and the second terms in $L_{eff}$ \rf{Lagr} averaged
        over the nucleon states give the spin-dependent and the
        spin-independent matrix elements ${\bf M}_{sd}$ and ${\bf M}_{si}$,
        respectively.
        For the spin-dependent amplitude we have \cite{Witt}, \cite{EF2}:
\be{Msd}  
   {\bf M}_{sd} = 4 \vec{S_\chi} \vec{S}_{p(n)}
                \sum_{q \in p(n)}^{} \{{\cal B}(q) - {\cal A}(q) \} \Delta q,
\ee
     where $\vec{S_{\chi}} \mbox{ and } \vec{S}_{p(n)}$ are
        the neutralino and proton (neutron) spin operators; $\Delta q$ are
        the fractions of the nucleon spin carried by the quark $q$.
        The standard definition is
\be{spin} 
   <p(n)|\bar q\gamma^\mu\gamma_5 q|p(n)> = 2 S_{p(n)}^{\mu} \Delta q,
\ee
      where $S_{p(n)}^{\mu}=(0,\vec{S}_{p(n)})$ is the 4-spin of the nucleon.
        The parameters $\Delta q$ (for the proton) can be extracted from
        the EMC \cite{EMC} and hyperon data \cite{Hyp}:
\be{spin_fract}  
        \Delta u =  0.77 \pm 0.08, \ \
        \Delta d = -0.49 \pm 0.08, \ \
        \Delta s = -0.15 \pm 0.08.
\ee
        The relevant values for the neutron can be found
        from \rf{spin_fract} by the isospin symmetry substitution
        $u\rightarrow d$, $d\rightarrow u$.

        The spin-independent matrix element
        has the form \cite{Gelm},
        \cite{FOT}\footnote{When this paper had been completed we received
        a paper ref.\cite{Drees} with more refined treatment of
        the spin-independent matrix element.}:

\ba{Msi} 
  {\bf M}_{si} &=& \left[ \hat f \frac{m_u {\cal C}(u)
        + m_d {\cal C}(d)}{m_u + m_d}
+ f {\cal C}(s)\right. \\
\nn            &+& \left.\frac{2}{27}(1-f - \hat f)({\cal C}(c)
        + {\cal C}(b) + {\cal C}(t))\right]\cdot
         \frac{M_{p(n)}}{M_{W}}\cdot\bar \chi\chi\cdot\bar \Psi \Psi,
\ea
         where the parameters $f$ and $\hat f$  are defined as follows:
\ba{Scal} 
   <p(n)|(m_{u} + m_{d})(\bar{u}u + \bar{d}d)|p(n)> &=& 2\hat f M_{p(n)}
           \bar \Psi \Psi, \\
\nn
   <p(n)|m_{s}\bar{s}s|p(n)> &=& f M_{p(n)}\bar \Psi \Psi.
\ea
        The values extracted from the data are \cite{Cheng},\cite{Gasser}:
        $\hat{f} = 0.05$ and $f = 0.2$.

        Averaging \rf{Msd}, \rf{Msi} over the nuclear states $|A>$ we deal
        with the following matrix elements at vanishing momentum transfer:
\ba{Nucl} 
   <A|M_{p(n)}\bar \Psi \Psi|A> &=& M_{A} \bar{A}A,  \\
\nn
   <A|\vec{S}_{p(n)}|A> &=& \lambda<A|\vec{J}|A>.
\ea
        Here $\vec{J}$ is the nuclear spin.
        On the basis of the odd-group shell model \cite{EV}
        (essentially somewhat relaxed single particle shell model)
        the parameter $\lambda$ can be related to the nuclear
        magnetic moment, $\mu$, as follows
\ba{lamb}   
   \lambda J = \frac{\mu - g^{l}J}{g^{s} - g^{l}},
\ea
        where $g^{l} = 1(0)$ and $g^{s} = 5.586(-3.826)$ are orbital and
        spin proton (neutron) $g$-factors.
        Then one can extract values of $\lambda$ for various nuclei from
        the experimental data on nuclear magnetic moments
        \footnote{A more direct way of calculation based on
        the theory of finite Fermi systems is presented in \cite{Nikolaev}.}.
        We use in this paper the values of $\lambda$ as presented
        in  ref.\cite{EV}.

        For large $M_\chi$ and $M_A$ the momentum transfer may be comparable
        to the inverse radius of a nucleus and then  we have to take into
        account the finite size effect (see also \cite{Nikolaev}).
        It can be done by introducing the coherence loss
        factor \cite{Gould}.
\be{Corr}      
        \zeta(r) = \frac{0.573}{b}
               \biggl(
            1 - \frac{\exp\bigl(-\frac{b}{1+b} \bigr)}{\sqrt{1+b}}
   \frac{ \mbox{erf} \bigl( \sqrt{ \frac{1}{1+b} } \bigr) }{\mbox{erf}(1)}
               \biggr),
\ee
        where
$$
     b = \frac{8}{9} \sigma^2 r^2
       \frac{ M_\chi^2 M^2_{A} }{ (M_\chi + M_{A})^2 }.
$$
        Here $\sigma^2$ is the dispersion of the Maxwellian neutralino
        velocity distribution $\sigma = 0.9 \cdot 10^{-3}$.
        To obtain the coherence loss factor for spin-independent
        scattering we take  $r = r_{charge}$ in \rf{Corr},
        where $r_{charge}$ is the $rms$ charge radius of the
        nucleus $A$ \cite{EF1}:
\be{rms} 
  r_{charge}=(0.3+0.89 M_{A}^{1/3})\mbox{ fm}.
\ee

        The coherence loss factor  for spin-dependent scattering is given by
        \rf{Corr} with $r = r_{spin}$.
        The $rms$ spin radius of the nucleus $A$ can be estimated as
        $r_{spin} = \xi\cdot r_{charge}$ with $\xi \approx 1.25$
        from harmonic well potential calculations \cite{EF1}.

        Finally we arrive at the formula for the event rate of elastic
        neutralino-nucleus scattering in the detector per day per unit mass
        of the target material:
\be{Rate} 
        R = R_{si} + R_{sd},
\ee
        where the spin-dependent and spin-independent parts are:
\ba{Rsd} 
   R_{sd} &=& 5.8\cdot 10^{10}\cdot \lambda^2 J(J+1) \zeta(r_{spin})
                {\cal M}_{sd}^2 \cdot {\cal D},\\
   R_{si} &=&  1.44\cdot 10^{10}\cdot \bigl(\frac{M_A}{M_W}\bigr)^2
 \zeta(r_{charge})\cdot {\cal M}_{si}^2 \cdot {\cal D}.
\ea

        The common kinematic factor $\cal D$ and properly normalized
        nucleon matrix elements ${\cal M}_{si},{\cal M}_{sd}$ are defined as:
\ba{D_Kin} 
    {\cal D} &=& \Bigl[ \frac{4M_\chi M_A}{(M_\chi + M_A)^2}  \Bigr]
                 \Bigl[ \frac{\rho}{ .3 GeV\cdot cm^{-3} }    \Bigr]
                 \Bigl[ \frac{<|\vec{v_E}|>}{ 320 km/s }      \Bigr]
                   \frac{ events }{  kg \cdot day }\\
{\bf M}_{si} &=& {\cal M}_{si}\cdot \frac{M_{p(n)}}{M_{W}}
                      \bar \chi\chi\cdot\bar \Psi \Psi,\\
{\bf M}_{sd} &=& 4\cdot {\cal M}_{sd}\cdot \vec{S}_{\chi}\vec{S}_{p(n)}.
\ea
        For the definition of ${\bf M}_{si}$, ${\bf M}_{sd}$ see formulae
        \rf{Msd}, \rf{Msi}.
        Here $\rho \approx$ 0.3 GeV$\cdot$cm$^{-3}$ is the DM
        neutralino density in the solar vicinity and
        $<|\vec{v_E}|> \approx$ 320 km/s is DM
        neutralino averaged velocity at the earth's surface.

        To study the role of nuclear spin in elastic $\chi$-nucleus
        scattering we introduce the ratio
\be{kappa} 
        \kappa = R_{sd}/R_{si}
\ee
        characterizing the relative contribution of spin-dependent and
        spin-independent interactions.
        From the practical point of view it determines the expected
        relative sensitivity of DM detectors with spin-non-zero
        (J $\neq$ 0) and spin-zero (J = 0) nuclei as target material.
        If $\kappa < 1$, then detectors with spin-non-zero and
        spin-zero target materials have approximately equal sensitivities
        to the DM signal, whereas if $\kappa > 1$ then, the spin-non-zero
        detectors are more sensitive than the spin-zero ones.

        Let us consider separately the dependence of $\kappa$ on the nuclear
        structure and the choice of a specific SUSY model.
        We may write:
\be{separat} 
   \kappa = \eta_{A}^{} \eta_{susy}^{p(n)},
\ee
where
\be{etaA} 
   \eta_{A}^{}= 4.03 \lambda^2 J(J+1)\cdot \frac{\zeta(r_{spin}) M_{W}^2}
               {\zeta(r_{charge}) M_{A}^2},
\ee
\be{eta} 
   \eta_{susy}^{p(n)}= \left(\frac{{\cal M}_{sd}^{p(n)}}
{{\cal M}_{si}}\right)^2.
\ee
        Here $\eta_{A}$ is a factor depending on the properties of the
        nucleus $A$; $\eta_{susy}^{p(n)}$ is defined by the SUSY-model which
        specifies the neutralino composition and the interactions with matter.
        The SUSY-factor also depends on the shell-model class
        to which nucleus $A$ belongs,
        being $\eta_{SUSY}^{n}$  for the shell-model "neutron"
        (${}^{3}$He, ${}^{29}$Si, ${}^{73}$Ge,...) and $\eta_{SUSY}^{p}$ for
        the shell-model "proton" (${}^{19}$F, ${}^{35}$Cl, ${}^{205}$Tl,...).

        Fig.1 depicts the nuclear factor $\eta_{A}$ versus
        the atomic weight $A$.
        The error bars represent the interval of the neutralino masses
        20 GeV$< M_{\chi} <$ 200 GeV.
        The lower bound corresponds to the present experimental constraints
        \cite{Roszk}-\cite{Part}.
        The upper bound is taken to include recent estimations for the mass
        of the cosmologically favorable neutralino \cite{Roberts}.
        It follows from  fig.1 that $\eta_{A} < 1$ for $A > 50$.
        Thus at $A > 50$ there is no nuclear structure enhancement of
        the spin-dependent event rate as compared to the spin-independent one.

        The next is an estimation of the SUSY-factor $\eta_{susy}^{p(n)}$.

\section{Specific SUSY-model Predictions.}

        To estimate the factor $\eta_{SUSY}$ in \rf{eta} one should calculate
        the parameters ${\cal A}(q)$, ${\cal B}(q)$ and ${\cal C}(q)$ of
        the effective Lagrangian \rf{Lagr} in the specific SUSY model.
        We will follow the MSSM.
        This model is specified by the superpotential and "soft"
        SUSY breaking terms \cite{Haber}.

        The effective low energy superpotential is:
\be{SUPERPOTENTIAL} 
         \tilde{W} = \sum_{generations}^{}
         (h_U \hat H_2 \hat Q \hat U + h_D \hat H_1 \hat Q\hat D +
        h_L \hat H_1 \hat L \hat E) + \mu \hat H_1\hat H_2.
\ee
        $H_1$ and $H_2$ are the Higgs fields with a weak hypercharge
        $Y=-1,+1$ respectively.

        SUSY breaking in the "hidden" sector of N=1 supergravity produces
        "soft" supersymmetry breaking terms in the scalar potential:
\ba{V_soft} 
V_{soft}= \sum_{i=scalars}^{}  m_{i}^{2} |\phi_i|^2 +
           h_U A_U H_2 \tilde Q \tilde U  + h_D A_D H_1 \tilde Q \tilde D +
h_L A_L H_1 \tilde L \tilde E && \\
\nn
+ \mu B H_1 H _2 + \mbox{ h.c.}  &&
\ea
         and a "soft"  gaugino mass term
\be{M_soft} 
{\cal L}_{FM}\  = \ - \frac{1}{2}\left[M_{1}^{} \tilde B \tilde B +
 M_{2}^{} \tilde W^3 \tilde W^3  + M_{3}^{} \tilde g^a \tilde g^a\right]
 -   \mbox{ h.c.}
\ee

        The  model is also characterized by the set of boundary conditions at
        the unification scale $M_X$:
\be{boundary1}  
 A_U = A_D = A_L = A_{0},
\ee
\be{boundary2}  
 m_{H_1}^{}=m_{H_2}^{}=m_{L}^{}=m_{E}^{}=m_{Q}^{}=m_{U}^{}=
  m_{D}^{}=m_{0},
\ee
\be{boundary3}  
 M_{1}^{} = M_{2}^{} = M_{3}^{} = m_{1/2}^{},
\ee
\be{} 
g_{1}^{}(M_X) = g_{2}^{}(M_X) = g_{2}^{}(M_X) =  g_{GUT}^{},
\ee
        where $g_{3}^{}, g_{2}^{}, g_{1}^{}$ are the $SU(3)\times SU(2)
        \times U(1)$ gauge coupling constants equal to $g_{GUT}^{}$  at the
        unification scale $M_X$.
        At the Fermi scale $Q\sim M_W$ these parameters
        can be evaluated on the basis of the 1-loop
        renormalization group equations (RGE) \cite{RGE1},\cite{RGE2}.

        The neutralino mass matrix in this model has the form \cite{Haber}:
\be{MassM}  
                M_{\chi} =  \left(
                        \begin{array}{cccc}
 M_2 & 0 & -M_Z c^{}_W s_\beta & M_Z c^{}_W c_\beta \\
 0 & M_2\frac{5}{3} t^{}_W
& M_Z s^{}_W s_\beta & -M_Z c^{}_W s_\beta\\
 -M_Z c^{}_W s_\beta &  M_Z s^{}_W s_\beta & 0 & -\mu  \\
  M_Z c^{}_W c_\beta & -M_Z s^{}_W c_\beta & -\mu & 0 \\
                        \end{array}
                     \right).
\ee
        where $c^{}_W = \cos\theta_W$, $s^{}_W = \sin\theta_W$,
        $t^{}_W = \tan \theta_W$, $s_\beta = \sin\beta$,
        $c_\beta = \cos\beta$.
        The matrix is written in the basis of fields $(\tilde{W}^{3}$,
        $\tilde{B}$, $\tilde{H}_{2}^{0}$, $\tilde{H}_{1}^{0})$.
        As usual, $M_2,\mu$ are the gaugino mass and the Higgs mixing
        parameter; the angle $\beta$ is defined by the vacuum expectation
        values of the neutral components of the Higgs fields:
        $\tan\beta = <H_{2}^{0}>/<H_{1}^{0}> = v_2/v_1$.
        Diagonalizing the mass matrix \rf{MassM} we obtain the lightest
        neutralino of the mass $M_\chi$ with the field content
$$
\chi = \alpha\tilde{W}^{3} + \beta\tilde{B} + \gamma\tilde{H}_{2}^{0} +
\delta\tilde{H}_{1}^{0}.
$$
        To calculate the low-energy neutralino-quark interactions we also
        need to have the spectrum of squarks $\tilde q$ and Higgs particles
        at the Fermi scale.
        Their mass matrices depend on the soft SUSY breaking parameters.
        We obtain them from the boundary conditions at the GUT scale
        $M_X$ \rf{boundary1}-\rf{boundary3} as a solution of the 1-loop RGE.

        We analyze the Higgs sector of the MSSM at the 1-loop
        level \cite{1loop}.
        In the analysis we take into account $\tilde{t}_{L}-\tilde{t}_{R}$,
        $\tilde{b}_{L}-\tilde{b}_{R}$ mixing between the third-generation
        squarks.
        Diagonalization of the Higgs mass matrix gives three neutral
        mass-eigenstates.
        There are two $CP$-even states, $H$, $h$, with the masses
        $m_H$, $m_h$ and the relevant mixing angle $\alpha_H$ and one
        $CP$-odd $A$ with the mass $m_A$.
        We take the mass $m_A$ as an independent free parameter of the MSSM.

        A complete list of essential free parameters of the MSSM is
\ba{param} 
\tan\beta, A_{0}, B, \mu, m_{1/2}, m_A, m_0, m_t.
\ea
        Having a particle spectrum one can derive the effective
        Lagrangian $L_{eff}$ of low-energy neutralino-quark interactions.
        As discussed in the previous section, its general form is given
        by eqn. \rf{Lagr}.
        In the MSSM the first term of $L_{eff}$ is induced by the $Z$-boson
        and $\tilde{q}$ exchange \cite{EHNOS} whereas the second one is
        due to the Higgs particle \cite{BFG} and $\tilde{q}$
        exchange \cite{Griest} as well as
        $\tilde{q}_{L}-\tilde{q}_{R}$ mixing \cite{Witt},\cite{SW}.
        The coefficients of $L_{eff}$ are
\begin{eqnarray}
\nn
 {\cal A}(q)&=&-\frac{|\gamma|^2-|\delta|^2}{4 M_Z^2}
              (g_1\sin\theta_W+g_2\cos\theta_W)
              \bigl(
              \frac{Y_L}{2} g_1\sin\theta_W - T_3 g_2 \cos\theta_W
              \bigr)\\
\label{(Aq)} 
&+&\frac{1}{2} \frac{|\alpha g_2 T_3 + \beta g_1 \frac{Y_L}{2}|^2}
                             {m^{2}_{\tilde{q}L} - M^2_\chi}
  + \frac{1}{2}\frac{m_q^2}{m^{2}_{\tilde{q}R} - M^2_\chi}
      \bigl[
          \frac{\frac{1}{2}+T_3}{v^2_2}|\gamma|^2
        + \frac{\frac{1}{2}-T_3}{v^2_1}|\delta|^2
                \bigr],
\end{eqnarray}
\ba{Bq} 
        {\cal B}(q) &=& -\frac{|\gamma|^2-|\delta|^2}{4 M_Z^2}
              (g_1\sin\theta_W+g_2\cos\theta_W)
              \bigl(
              \frac{Y_R}{2} g_1\sin\theta_W
              \bigr) \mbox{ \hspace*{2cm} } \\
\nn      &-&\frac{1}{2} \frac{|\beta g_1 \frac{Y_R}{2}|^2}
                   {m^{2}_{\tilde{q}R} - M^2_\chi}
        - \frac{1}{2}\frac{m_q^2}{m^{2}_{\tilde{q}L} - M^2_\chi}
        \bigl[
  \frac{\frac{1}{2}+T_3}{v^2_2}|\gamma|^2
+ \frac{\frac{1}{2}-T_3}{v^2_1}|\delta|^2
                \bigr],
\ea

\ba{Cq}   
      {\cal C}(q) &=& \frac{g_2^2}{4 m^2_{h1}} F_q
        \bigl[ (\frac{1}{2}+T_3)\frac{\cos\alpha_H}{\sin\beta}
             - (\frac{1}{2}-T_3)\frac{\sin\alpha_H}{\cos\beta}
        \bigr]  \\
\nn      &+& \frac{g_2}{4}
        \bigl[
       \frac{\alpha g_2 T_3+\frac{Y_L}{2}\beta g_1}{m^{2}_{\tilde{q}L}
                                - M^2_\chi}
      -\frac{\frac{Y_R}{2}g_1\beta}{m^{2}_{\tilde{q}R} - M^2_\chi}
        \bigr]
        \bigl[ (\frac{1}{2}+T_3) \frac{\gamma}{\sin\beta}
             - (\frac{1}{2}-T_3) \frac{\delta}{\cos\beta}
        \bigr].
\ea
     Here
\be{Fq}  
        F_q = (\alpha - \beta\tan\theta_W)
        (\gamma\cos\alpha_H + \delta\sin\alpha_H).
\ee
        In these formulae we ignore $\tilde{q}_{L}-\tilde{q}_{R}$ mixing
        since they give a small contribution according to the estimation
        of ref.~\cite{EF1}.

        The procedure we use for the neutralino mass matrix diagonalization
        always leads to positive mass eigenvalues and to either real
        or pure imaginary values of the coefficients
        $\alpha,\beta,\gamma,\delta$.
        Therefore in formulae \rf{Aq},\rf{Bq} the absolute values
        of these coefficients appear.

        Now we are ready to calculate the $\eta_{susy}$-factor \rf{eta}
        substituting the definitions \rf{Aq}--\rf{Cq} in formula \rf{eta}.

        To get complete information about possible values of the
        $\eta_{susy}$-factor we scan the MSSM parameter space within
        the constraints imposed by the experimental data and some
        general theoretical principles.
        The well known experimental constraints \cite{Part}
        are summarized in the Table.

\begin{table}[ht]
\centerline{\mbox{\bf Table:~Present Limits on Supersymmetric Particles.}}

\centerline{ \mbox{(Table is taken from ref.\cite{Haber1}.)} }
\vspace*{0.1cm}
\begin{tabular}{|c|c|cp{3.0in}|}   \hline
            & Bound on        & &       \\[-1mm]
Particle    & Particle Mass   & & \hspace*{2.2cm} Source  \\[-1mm]
            &  (GeV)          & &        \\ \hline
            &                 &  &    \\[-5mm]
\parbox[t]{0in}{ ${\tilde{\chi}} \, ^\circ_1$
              \newline ${\tilde{\chi}} \, ^\circ_2$
              \newline ${\tilde{\chi}} \, ^\circ_3$
              \newline ${\tilde{\chi}} \, ^\circ_4$ }
& \parbox[t]{0in}{18.4
\newline       45
\newline       70
\newline       108 }
&  ${}_{\Biggl\{ } \Biggr.$
& \parbox[t]{3.0in}{ Based on the LEP non--observation of
  ${\tilde{\chi}} \, ^\circ_i$ and ${\tilde{\chi}} \, ^+_1$,
        CDF non--observation
 of $\tilde{g}$, and the assumption of gaugino mass unification.}   \\[2mm]
\hspace*{1mm}$\quad {\tilde{\chi}} \, ^\pm_1$   & $\qquad 45.2$  &  & LEP\\
\hspace*{1mm}$\quad {\tilde{\chi}} \, ^\pm_2$   & $\quad 99$     &  &
See neutralino mass limits above.  \\
$\quad {\tilde{\nu}}$               & $\quad 41$     &  &
\parbox[t]{3.0in}{Assumes the $\tilde{\nu}$ decays are
  invisible (otherwise $M \, _{\tilde{\nu}} \, < \, 32$~GeV). Based on LEP
 measurement of $\Gamma (Z \, \rightarrow \, $ invisible final states).
        } \\[2mm]
$\quad {\tilde{e}}$    & $\quad 45$&  & LEP;
        assumes $M \, _{{\tilde{\chi}} \, ^\circ_1} \, < 41$~GeV        \\
$\quad {\tilde{\mu}}$  & $\quad 45$&  & LEP;
        assumes $M \, _{{\tilde{\chi}} \, ^\circ_1} \, < 41$~GeV        \\
$\quad {\tilde{r}}$    & $\quad 45$&  & LEP;
        assumes $M \, _{{\tilde{\chi}} \, ^\circ_1} \, < 38$~GeV        \\
$\quad {\tilde{q}}$    & $\quad 45$&  & LEP;
        assumes $M \, _{{\tilde{\chi}} \, ^\circ_1} \, < 20$~GeV        \\
                       & $\quad 74$&  & UA2 (any $M \, _{\tilde{g}}$)   \\
                       & $\sim 95$ &  & CDF
        ($M \, _{\tilde{q}} \, < \, M \, _{\tilde{g}}$)  \\
$\quad {\tilde{g}}$    & $\quad 79$&  & UA2
        ($M \, _{\tilde{g}} \, < \, M \, _{\tilde{q}}$)  \\
                       & $\sim 95$ &  & CDF   \\[-2mm]
                       &   &  &  \\ \hline
\end{tabular}
\end{table}
        The constraints are given for masses of squarks $\tilde q$,
        the gluino $\tilde g$, charginos $\chi^{\pm}$, the neutralino
        $\chi$, charged sleptons $\tilde l^{\pm}$, the sneutrino
        $\tilde\nu$, lightest $CP$-even $h$ and $CP$-odd $A$ Higgs bosons.
        We also include the constraints
\be{tan} 
1.12 < \tan\beta < 4.7.
\ee
        The lower limit follows from the finiteness condition for the top
        Yukawa coupling $Y_t$.
        If
\be{finitn} 
 \sin\beta > (m_t/200~GeV),
\ee
        then $Y_t$ is finite up to the unification scale $M_X$.
        For $m_t = 150$~GeV we get the lower limit in \rf{tan}.
        The upper limit in this formula is expected from proton
        stability considerations \cite{Arnowitt}.

        From the "naturalness" arguments \cite{BG} we may choose:
\be{sparticles}  
        m_{\tilde q}, m_{\tilde g} < 1~TeV,
\ee
        where $m_{\tilde f}$ is the mass of any sfermion $\tilde f$.
        The choice of the interval for the neutralino mass
$$
        20~GeV < M_{\chi} < 200~GeV
$$
        was already explained at the end of {\it section 2}.

        The additional constraint we use in the analysis of the role
        of nuclear spin is the constraint on the realistic sensitivity of
        the DM detector.
        In terms of the total event rate $R$ we choose the sensitivity to be
        not better than:
\be{sens} 
R > 0.01 \frac{event}{kg \cdot days}.
\ee
        We do not expect the DM detectors to go
        below this lower bound in near future.
        Therefore the constraint \rf{sens} reflects the realistic capacities
        of the present and near-future set-ups.
        It excludes the region in the parameter space corresponding to
        the low-level DM signals inaccessible to these detectors.

        We have performed a numerical analysis of the MSSM parameter space
        within the above-defined constraints.
        In fig.2  the typical behavior of the $\eta_{susy}$-factor in
        particular domains of the MSSM parameter space is presented.
        The following upper bound for the SUSY-factor in
        eqn. \rf{separat} was found:
\be{main}  
\eta_{susy} \leq 1.2.
\ee
        Combining this result with the values of the nuclear factor
        $\eta_A^{}$ represented in fig.1 we conclude that
\be{main1} 
  \kappa = R_{sd}/R_{si} =
        \eta_A^{} \eta_{susy}^{p(n)} \leq 1 \
        \mbox{ ~for nuclei with~} A > 50
\ee
        at a detector sensitivity up to $R > 0.01$.
        The tendency is that at higher sensitivities (lower $R$ accessible)
        we get $\kappa \leq 1$ for heavier nuclei.

        As a by-product of our analysis in fig.3 we also give
        the event rate for some nuclei of special interest in DM search.

        We do not take into account possible rescaling of the local
        neutralino density $\rho$ which may occur in the region of the
        MSSM parameter space where $\Omega h^2<0.05$ \cite{Gelm}.
        This effect, if it took place, would essentially reduce the
        event rate $R$ \cite{Bot1}.
        Of course, it has no influence on the ratio $\kappa$ in the
        formula \rf{kappa} and on our conclusion about the role
        of nuclear spin.
        Plots in fig.3 correspond to a situation when neutralinos
        constitute a dominant component of the DM halo of our galaxy
        with the density $\rho$ = 0.3 GeV$\cdot$cm$^{-3}$ in the
        solar vicinity.

\section{Conclusions}

        In the framework of general assumptions about the nuclear and
        nucleon structure considering the MSSM as the basis for description
        of the neutralino properties we have drawn the following
        basic conclusions.

        For sufficiently heavy nuclei with atomic weights $A > 50$
        the spin-independ\-ent event rate $R_{si}$ is larger than
        the spin-dependent one $R_{sd}$ if low-level signals
        with total event rates $R =R_{sd} + R_{si} < 0.01$ are ignored.
        This cut condition reflects the realistic sensitivities of
        the present and the near-future DM detectors.

        The main practical issue is that two different DM detectors with
        (J~=~0, A$_1$) and with (J~$\neq$~0, A$_2$) nuclei as target
        material have equal chances  to discover the DM event
        if  A$_1\sim$ A$_2 >$ 50.

        Another aspect of  the DM search is the investigation of the
        SUSY-model parameter space  from nonobservation of DM events.
        Apparently, in this case experiments with J~$\neq$~0 nuclei are
        important since they provide new information about the SUSY model
        parameters from  $R_{sd}$ which is inaccessible in
        J~=~0 experiments.

        The  results presented above were obtained in a specific SUSY-model.
        Therefore it is a natural question whether our basic conclusions
        hold for other popular SUSY-models. We plan to investigate
        this question in a subsequent paper.


\newpage

\bigskip

{\large\bf Figure Caption}\\

\begin{itemize}
\item[Fig.1]  The nuclear factor $\eta_{A}$ versus
        the atomic weight $A$.
        The error bars represent the interval
        of the neutralino masses 20~GeV$ < M_{\chi} <$ 200~GeV.

\item[Fig.2]  The SUSY factor $\eta_{susy}$
        versus neutralino mass $M_{\chi}$
        at various values of the MSSM free parameters. (a) and (b)
        for nuclei with proton and neutron shell model structure,
        respectively.

\item[Fig.3]  Samples of plots for the total event rate for
                neutralino elastic scattering off
                (a) equal parts mixture of $^{73}Ge+^{76}Ge$ and
                (b) $^{19}F$ nuclei.
\end{itemize}

\end{document}